\documentclass[12pt]{article}

\makeatletter

\@addtoreset{equation}{section}
\makeatother

\newtheorem{definition}{Definition}[section]
\newtheorem{theorem}{Theorem}[section]

\newtheorem{lemma}{Lemma}[section]
\newtheorem{remark}{Remark}[section]

\usepackage{amssymb}
\usepackage{latexsym}

\begin{document} 

\title{Uncertainty Relation on Wigner-Yanase-Dyson Skew Information}
\author{Kenjiro Yanagi\thanks{
Division of Applied Mathematical Science,
Graduate School of Science and Engineering,
Yamaguchi University,
Ube, 755-8611 Japan. 
E-mail: {\tt yanagi@\allowbreak
yamaguchi-u.\allowbreak
ac.\allowbreak
jp},  
This research was partially supported by the Ministry of Education, Science, Sports and Culture, Grant-in-Aid for 
Scientific Research (B), 18300003 and (C), 20540175}}
\date{}

\maketitle

\begin{flushleft}
{\bf Abstract.} We give a trace inequality related to the uncertainty relation 
of Wigner-Yanase-Dyson skew information. This inequality corresponds to a generalization 
of the uncertainty relation derived by S.Luo \cite{Lu:Hei} for the quantum uncertainty quantity 
excluding the classical mixture.
\end{flushleft}

\begin{flushleft}
\end{flushleft}

\section{Introduction}

Wigner-Yanase skew information
\begin{eqnarray*}
I_{\rho}(H) & = & \frac{1}{2}Tr \left[ \left( i \left[ \rho^{1/2},H \right] \right)^2 \right] \\
& = & Tr[\rho H^2]-Tr[\rho^{1/2}H\rho^{1/2}H] 
\end{eqnarray*}
was defined in \cite{WY:inf}. This quantity can be considered as a kind of 
the degree for non-commutativity between a quantum state $\rho$ and an 
observable $H$. Here we denote the commutator by $[X,Y] = XY-YX$. This 
quantity was generalized by Dyson 
\begin{eqnarray*}
I_{\rho,\alpha}(H) & = & \frac{1}{2}Tr[(i[\rho^{\alpha},H])(i[\rho^{1-\alpha},H])] \\
& = & Tr[\rho H^2]-Tr[\rho^{\alpha}H\rho^{1-\alpha}H],  \alpha \in [0,1]
\end{eqnarray*}
which is known as the Wigner-Yanase-Dyson skew information. It is famous that 
the convexity of $I_{\rho,\alpha}(H)$ with respect to $\rho$ was successfully 
proven by E.H.Lieb in \cite{Li:convex}. From the physical point of view, 
an observable $H$ is generally considered to be an unbounded opetrator, 
however in the present paper, unless otherwise stated, we consider 
$H \in B({\cal H})$ represents the set of all bounded linear operators on 
the Hilbert space ${\cal H}$, as a mathematical interest. We also denote 
the set of all self-adjoint operators (observables) by ${\cal L}_h({\cal H})$ 
and the set of all density operators (quantum states) by ${\cal S}({\cal H})$ 
on the Hilbert space ${\cal H}$. The relation between the Wigner-Yanase skew 
information and the uncertainty relation was studied in \cite{LuZh:skew}. 
Moreover the relation between the Wigner-Yanase-Dyson skew information and the 
uncertainty relation was studied in \cite{Ko:matrix, YaFuKu:gen}. In our paper 
\cite{YaFuKu:gen}, we defined a generalized skew information and then derived 
a kind of an uncertainty relation. In the section 2, we discuss various 
properties of the Wigner-Yanase-Dyson skew information. Finally in section3, 
we give our main result and its proof.

\section{Trace inequalities of Wigner-Yanase-Dyson skew information}

We review the relation between the Wigner-Yanase skew information and the uncertainty relation. 
In quantum mechanical system, the expectation value of an observable $H$ 
in a quantum state $\rho$ is expressed by $Tr[\rho H]$. It is natural 
that the variance for a quantum state $\rho$ and an observable $H$ is defined by 
$V_{\rho}(H) = Tr[\rho (H-Tr[\rho H]I)^2] = Tr[\rho H^2]-Tr[\rho H]^2$. 
It is famous that we have 
\begin{equation}
V_{\rho}(A)V_{\rho}(B) \geq \frac{1}{4}|Tr[\rho [A,B]]|^2 
\label{eq:num2-1}
\end{equation}
for a quantum state $\rho$ and two observables $A$ and $B$. The further strong results 
was given by Schrodinger 
$$
V_{\rho}(A)V_{\rho}(B)-|Cov_{\rho}(A,B)|^2 \geq \frac{1}{4}|Tr[\rho [A,B]]|^2, 
$$
where the covariance is defined by 
$Cov_{\rho}(A,B) = Tr[\rho (A-Tr[\rho A]I)(B-Tr[\rho B]I)]$. 
However, the uncertainty relation for the Wigner-Yanase skew information failed. 
(See \cite{LuZh:skew, Ko:matrix, YaFuKu:gen}) 
$$
I_{\rho}(A) I_{\rho}(B) \geq \frac{1}{4}|Tr[\rho [A,B]]|^2.
$$
Recently, S.Luo introduced the quantity $U_{\rho}(H)$ representing a quantum uncertainty 
excluding the classical mixture: 
\begin{equation}
U_{\rho}(H) = \sqrt{V_{\rho}(H)^2-(V_{\rho}(H)-I_{\rho}(H))^2}, 
\label{eq:num2-2}
\end{equation}
then he derived the uncertainty relation on $U_{\rho}(H)$ in \cite{Lu:Hei}: 
\begin{equation}
U_{\rho}(A) U_{\rho}(B) \geq \frac{1}{4}|Tr[\rho [A,B]]|^2.
\label{eq:num2-3}
\end{equation}
Note that we have the following relation 
\begin{equation}
0 \leq I_{\rho}(H) \leq U_{\rho}(H) \leq V_{\rho}(H). 
\label{eq:num2-4}
\end{equation}
The inequality (\ref{eq:num2-3}) is a refinement of the inequality (\ref{eq:num2-1}) 
in the sense of (\ref{eq:num2-4}). 
In this section, we study one-parameter extended inequality for the inequality (\ref{eq:num2-3}). 

\begin{definition}
For $0 \leq \alpha \leq 1$, a quantum state $\rho$ and an observable $H$, we define 
the Wigner-Yanase-Dyson skew information 
\begin{eqnarray}
I_{\rho,\alpha}(H) & = & \frac{1}{2}Tr[(i [\rho^{\alpha},H_0])(i [\rho^{1-\alpha},H_0])] \nonumber \\
& = & Tr[\rho H_0^2]-Tr[\rho^{\alpha}H_0\rho^{1-\alpha}H_0] 
\label{eq:num2-5}
\end{eqnarray}
and we also define 
\begin{eqnarray}
J_{\rho,\alpha}(H) & = & \frac{1}{2}Tr[\{ \rho^{\alpha},H_0 \} \{ \rho^{1-\alpha},H_0 \}]  \nonumber \\
& = & Tr[\rho H_0^2]+Tr[\rho^{\alpha}H_0 \rho^{1-\alpha}H_0], 
\label{eq:num2-6}
\end{eqnarray}
where $H_0 = H-Tr[\rho H]I$ and we denote the anti-commutator by $\{ X,Y \} = XY+YX$.  
\label{def:definition2-1}
\end{definition}

Note that we have 
$$
\frac{1}{2}Tr[(i [\rho^{\alpha},H_0])(i [\rho^{1-\alpha},H_0])] = \frac{1}{2}Tr[(i [\rho^{\alpha},H])(i [\rho^{1-\alpha},H])]
$$
but we have 
$$
\frac{1}{2}Tr[\{ \rho^{\alpha},H_0 \} \{ \rho^{1-\alpha},H_0 \}] \neq \frac{1}{2}Tr[\{\rho^{\alpha},H \} \{ \rho^{1-\alpha},H \}]. 
$$
Then we have the following inequalities: 
\begin{equation}
I_{\rho,\alpha}(H) \leq I_{\rho}(H) \leq J_{\rho}(H) \leq J_{\rho,\alpha}(H), 
\label{eq:num2-7}
\end{equation}
since we have $Tr[\rho^{1/2}H\rho^{1/2}H] \leq Tr[\rho^{\alpha}H\rho^{1-\alpha}H]$. 
(See \cite{Bo:some,Fu:trace} for example.) If we define 
\begin{equation}
U_{\rho,\alpha}(H) = \sqrt{V_{\rho}(H)^2-(V_{\rho}(H)-I_{\rho,\alpha}(H))^2}, 
\label{eq:num2-8}
\end{equation}
as a direct generalization of Eq.(\ref{eq:num2-2}), then we have 
\begin{equation}
0 \leq I_{\rho,\alpha}(H) \leq U_{\rho,\alpha}(H) \leq U_{\rho}(H) 
\label{eq:num2-9}
\end{equation}
due to the first inequality of (\ref{eq:num2-7}). We also have 
$$
U_{\rho,\alpha}(H) = \sqrt{I_{\rho,\alpha}(H) J_{\rho,\alpha}(H)}. 
$$
From the inequalities (\ref{eq:num2-4}),(\ref{eq:num2-8}),(\ref{eq:num2-9}), our situation is that we have 
$$
0 \leq I_{\rho,\alpha}(H) \leq I_{\rho}(H) \leq U_{\rho}(H)
$$
and
$$
0 \leq I_{\rho,\alpha}(H) \leq U_{\rho,\alpha}(H) \leq U_{\rho}(H).
$$
Our concern is to show an uncertainty relation with respect to $U_{\rho,\alpha}(H)$ as a direct 
generalization of the inequality (\ref{eq:num2-3}) such that 
\begin{equation}
U_{\rho,\alpha}(A) U_{\rho,\alpha}(B) \geq \frac{1}{4}|Tr[\rho [A,B]]|^2 
\label{eq:num2-10}
\end{equation}

On the other hand, we introduced a generalized Wigner-Yanase skew information which is a generalization of 
the inequality (\ref{eq:num2-10}), but different from the Wigner-Yanase-Dyson skew information 
defined in (\ref{eq:num2-5}) and gave the following theorem in \cite{FuYaKu:trace}. 

\begin{theorem}
For $0 \leq \alpha \leq 1$, a quantum state $\rho$ and an observable $H$, 
we define a generalized Wigner-Yanase skew information by 
$$
K_{\rho,\alpha}(H) = \frac{1}{2}Tr \left[ \left( i \left[ \frac{\rho^{\alpha}+\rho^{1-\alpha}}{2}, H_0 \right] \right)^2 \right]
$$
and we also define 
$$
L_{\rho,\alpha}(H) = \frac{1}{2}Tr \left[ \left( i \left\{ \frac{\rho^{\alpha}+\rho^{1-\alpha}}{2}, H_0 \right\} \right)^2 \right], 
$$
and
$$
W_{\rho,\alpha}(H) = \sqrt{K_{\rho,\alpha}(H)L_{\rho,\alpha}(H)}.
$$
Then for a quantum state $\rho$ and observables $A, B$ and $\alpha \in [0,1]$, we have 
$$
W_{\rho,\alpha}(A)W_{\rho,\alpha}(B) \geq \frac{1}{4}\left|Tr \left[ \left( \frac{\rho^{\alpha}+\rho^{1-\alpha}}{2} \right)^2[A,B] \right] \right|^2.
$$
\label{th:theorem2-1}
\end{theorem}

\section{Main Theorem}

We give the main theorem as follows; 

\begin{theorem}
For a quantum state $\rho$ and observables $A, B$ and $0 \leq \alpha \leq 1$, we have 
\begin{equation}
U_{\rho,\alpha}(A)U_{\rho,\alpha}(B) \geq \alpha(1-\alpha)|Tr[\rho [A,B]]|^2.
\label{eq:num3-1}
\end{equation}
\label{th:theorem3-1}
\end{theorem}

We use the several lemmas to prove the theorem \ref{th:theorem3-1}. By spectral decomposition, there exists an orthonormal basis 
$\{ \phi_1, \phi_2,\ldots \}$ consisting of eigenvectors of $\rho$. Let $\lambda_1, \lambda_2,\ldots$ 
be the corresponding eigenvalues, where $\sum_{i=1}^{\infty} \lambda_i = 1$ and $\lambda_i \geq 0$. 
Thus, $\rho$ has a spectral representation 
\begin{equation}
\rho = \sum_{i=1}^{\infty} \lambda_i |\phi_i \rangle \langle \phi_i|.
\label{eq:num3-2}
\end{equation}

\begin{lemma}
$$
I_{\rho,\alpha}(H) = \sum_{i<j}(\lambda_i+\lambda_j-\lambda_i^{\alpha}\lambda_j^{1-\alpha}-\lambda_i^{1-\alpha}\lambda_j^{\alpha})|\langle \phi_i|H_0|\phi_j \rangle|^2.
$$
\label{lem:lemma3-1}
\end{lemma}

\begin{flushleft}
{\bf Proof of Lemma \ref{lem:lemma3-1}.} By (\ref{eq:num3-2}), 
\end{flushleft}
$$
\rho H_0^2 = \sum_{i=1}^{\infty} \lambda_i|\phi_i\rangle \langle \phi_i|H_0^2.
$$
Then 
\begin{equation}
Tr[\rho H_0^2]  =  \sum_{i=1}^{\infty} \lambda_i \langle \phi_i|H_0^2|\phi_i \rangle = \sum_{i=1}^{\infty} \lambda_i \| H_0|\phi_i \rangle \|^2. 
\label{eq:num3-3}
\end{equation}
Since 
$$
\rho^{\alpha}H_0 = \sum_{i=1}^{\infty} \lambda_i^{\alpha} |\phi_i \rangle \langle \phi_i|H_0
$$
and 
$$
\rho^{1-\alpha}H_0 = \sum_{i=1}^{\infty} \lambda_i^{1-\alpha} |\phi_i \rangle \langle \phi_i|H_0,
$$
we have 
$$
\rho^{\alpha}H_0 \rho^{1-\alpha}H_0 = \sum_{i,j=1}^{\infty} \lambda_i^{\alpha}\lambda_j^{1-\alpha}|\phi_i \rangle  
\langle \phi_i|H_0|\phi_j \rangle \langle \phi_j|H_0. 
$$
Thus 
\begin{eqnarray}
Tr[\rho^{\alpha}H_0 \rho^{1-\alpha}H_0] & = & \sum_{i,j=1}^{\infty} \lambda_i^{\alpha}\lambda_j^{1-\alpha} \langle \phi_i|H_0|\phi_j \rangle \langle \phi_j|H_0|\phi_i \rangle \nonumber \\
& = & \sum_{i,j=1}^{\infty} \lambda_i^{\alpha}\lambda_j^{1-\alpha} | \langle \phi_i|H_0|\phi_j \rangle |^2. \label{eq:num3-4}
\end{eqnarray}
From (\ref{eq:num2-5}), (\ref{eq:num3-3}), (\ref{eq:num3-4}), 
\begin{eqnarray*}
I_{\rho,\alpha}(H) & = & \sum_{i=1}^{\infty} \lambda_i \| H_0|\phi_i \rangle \|^2-\sum_{i,j=1}^{\infty} \lambda_i^{\alpha}\lambda_j^{1-\alpha}| \langle \phi_i|H_0|\phi_j \rangle |^2 \\
& = & \sum_{i,j=1}^{\infty} (\lambda_i-\lambda_i^{\alpha}\lambda_j^{1-\alpha})| \langle \phi_i|H_0|\phi_j \rangle |^2 \\
& = & \sum_{i<j} (\lambda_i+\lambda_j-\lambda_i^{\alpha}\lambda_j^{1-\alpha}-\lambda_i^{1-\alpha}\lambda_j^{\alpha})| \langle \phi_i|H_0|\phi_j \rangle |^2. 
\end{eqnarray*}
\ \hfill $\Box$

\begin{lemma}
$$
J_{\rho,\alpha}(H) \geq \sum_{i<j} (\lambda_i+\lambda_j+\lambda_i^{\alpha}\lambda_j^{1-\alpha}+\lambda_i^{1-\alpha}\lambda_j^{\alpha})| \langle \phi_i|H_0|\phi_j \rangle |^2. 
$$
\label{lem:lemma3-2}
\end{lemma}

\begin{flushleft}
{\bf Proof of Lemma \ref{lem:lemma3-2}.}  By (\ref{eq:num2-6}), (\ref{eq:num3-3}), (\ref{eq:num3-4}), we have 
\end{flushleft}
\begin{eqnarray*}
J_{\rho,\alpha}(H) & = & \sum_{i=1}^{\infty} \lambda_i \| H_0|\phi_i \rangle \|^2 + \sum_{i,j=1}^{\infty} \lambda_i^{\alpha}\lambda_j^{1-\alpha} |\langle \phi_i|H_0|\phi_j \rangle |^2 \\
& = & \sum_{i,j=1}^{\infty} (\lambda_i + \lambda_i^{\alpha}\lambda_j^{1-\alpha})|\langle \phi_i|H_0|\phi_j \rangle |^2 \\
& = & 2 \sum_{i=1}^{\infty} \lambda_i |\langle \phi_i|H_0|\phi_i \rangle |^2 + \sum_{i \neq j} (\lambda_i + \lambda_i^{\alpha}\lambda_j^{1-\alpha})| \langle \phi_i|H_0|\phi_j \rangle |^2 \\ 
& = & 2 \sum_{i=1}^{\infty} \lambda_i |\langle \phi_i|H_0|\phi_i \rangle |^2 + \sum_{i<j} (\lambda_i+\lambda_j+\lambda_i^{\alpha}\lambda_j^{1-\alpha}+\lambda_i^{1-\alpha}\lambda_j^{\alpha})| \langle \phi_i|H_0|\phi_j \rangle |^2 \\
& \geq & \sum_{i<j} (\lambda_i+\lambda_j+\lambda_i^{\alpha}\lambda_j^{1-\alpha}+\lambda_i^{1-\alpha}\lambda_j^{\alpha})| \langle \phi_i|H_0|\phi_j \rangle |^2. 
\end{eqnarray*}
\ \hfill $\Box$

\begin{lemma}
For any $t > 0$ and $0 \leq \alpha \leq 1$, the following inequality holds; 
\begin{equation}
(1-2 \alpha)^2(t-1)^2-(t^{\alpha}-t^{1-\alpha})^2 \geq 0.
\label{eq:num3-5} 
\end{equation}
\label{lem:lemma3-3}
\end{lemma}

\vspace{1cm}
\noindent
{\bf Proof of Lemma \ref{lem:lemma3-3}.}  If $\alpha = 0$ or $\frac{1}{2}$ or $1$, then it is clear that 
(\ref{eq:num3-5}) is satisfied.  Now we put 
$$
F(t) = (1-2 \alpha)^2(t-1)^2-(t^{\alpha}-t^{1-\alpha})^2.
$$
We have 
$$
F^{'}(t) = 2(1-2 \alpha)^2 t -2 \alpha t^{2 \alpha -1} - 2(1-\alpha)t^{1-2 \alpha} + 8 \alpha (1-\alpha).
$$
And we also have 
$$
F^{''}(t) = 2(1-2 \alpha)^2 - 2\alpha (2 \alpha -1) t^{2 \alpha -2} - 2(1-\alpha)(1-2 \alpha) t^{-2\alpha}
$$
and 
\begin{eqnarray*}
&   & F^{'''}(t) \\
& = & 4 \alpha (1-2 \alpha) (1-\alpha) t^{-2\alpha -1} - 4 \alpha (1-2 \alpha)(1-\alpha) t^{2\alpha -3} \\
& = & 4 \alpha (1-2\alpha)(1-\alpha) \left( \frac{1}{t^{1+2\alpha}}-\frac{1}{t^{3-2\alpha}} \right).
\end{eqnarray*}
If $\frac{1}{2} < \alpha < 1$, then $1+2\alpha > 3-2 \alpha$. Then it is easy to show that $F^{'''}(t) < 0$ for $t < 1$ 
and $F^{'''}(t) > 0$ for $t > 1$. On the other hand if $0 < \alpha < \frac{1}{2}$, then $1+2\alpha < 3-2 \alpha$. 
Then it is easy to show that $F^{'''}(t) < 0$ for $t < 1$ and $F^{'''}(t) > 0$ for $t > 1$. Since $F^{''}(1) = 0$, 
we can get $F^{''}(t) > 0$. Since $F^{'}(1) = 0$, we also have 
$F^{'}(t) < 0$ for $t < 1$ and $F^{'}(t) > 0$ for $t > 1$. Since $F(1) = 0$, we finally get 
$F(t) \geq 0$ for all $t > 0$. Therefore we have (\ref{eq:num3-5}). \ \hfill $\Box$ 

\vspace{1cm}
\noindent
{\bf Proof of Theorem \ref{th:theorem3-1}.} We put $\displaystyle{t = \frac{\lambda_i}{\lambda_j}}$ 
in (\ref{eq:num3-5}).  Then we have  
$$
(1-2\alpha)^2 \left( \frac{\lambda_i}{\lambda_j}-1 \right)^2 - \left( \left( \frac{\lambda_i}{\lambda_j} \right)^{\alpha} 
- \left( \frac{\lambda_i}{\lambda_j} \right)^{1-\alpha} \right)^2 \geq 0.
$$
And we get 
$$
(1-2 \alpha)^2 (\lambda_i -\lambda_j)^2 - (\lambda_i^{\alpha} \lambda_j^{1-\alpha} - \lambda_i^{1-\alpha} \lambda_j^{\alpha})^2 \geq 0
$$
and 
$$
(\lambda_i -\lambda_j)^2 - (\lambda_i^{\alpha}\lambda_j^{1-\alpha}-\lambda_i^{1-\alpha}\lambda_j^{\alpha} )^2 \geq 4 \alpha(1-\alpha)(\lambda_i - \lambda_j)^2
$$
and
\begin{equation}
(\lambda_i +\lambda_j)^2 - (\lambda_i^{\alpha}\lambda_j^{1-\alpha}+\lambda_i^{1-\alpha}\lambda_j^{\alpha} )^2 \geq 4 \alpha(1-\alpha)(\lambda_i - \lambda_j)^2. 
\label{eq:num3-6}
\end{equation}
Since 
\begin{eqnarray*}
Tr[\rho[A,B]] & = & Tr[\rho[A_0,B_0]] \\
& = & 2i Im Tr[\rho A_0B_0] \\ 
& = & 2i Im \sum_{i<j} (\lambda_i-\lambda_j) \langle \phi_i|A_0|\phi_j \rangle \langle \phi_j|B_0|\phi_i \rangle \\
& = & 2i \sum_{i<j} (\lambda_i-\lambda_j) Im \langle \phi_i|A_0|\phi_j \rangle \langle \phi_j|B_0|\phi_i \rangle, 
\end{eqnarray*}
\begin{eqnarray*}
|Tr[\rho[A,B]]| & = & 2|\sum_{i<j} (\lambda_i-\lambda_j) Im \langle \phi_i|A_0|\phi_j \rangle \langle \phi_j|B_0|\phi_i \rangle | \\
& \leq & 2 \sum_{i<j} |\lambda_i-\lambda_j||Im \langle \phi_i|A_0|\phi_j \rangle \langle \phi_j|B_0|\phi_i \rangle |. 
\end{eqnarray*}
Then we have 
$$
|Tr[\rho[A,B]]|^2 \leq 4 \{ \sum_{i<j} |\lambda_i-\lambda_j|| Im \langle \phi_i|A_0|\phi_j \rangle \langle \phi_j|_0|\phi_i \rangle | \}^2.
$$
By (\ref{eq:num3-6}) and Schwarz inequality, 
\begin{eqnarray*}
&   & \alpha(1-\alpha)|Tr[\rho[A,B]]|^2 \\
& \leq & 4\alpha(1-\alpha) \{\sum_{i<j} |\lambda_i-\lambda_j|| Im \langle \phi_i|A_0|\phi_j \rangle \langle \phi_j|B_0|\phi_i \rangle | \}^2 \\
& = & \{ \sum_{i<j} 2 \sqrt{\alpha(1-\alpha)}|\lambda_i-\lambda_j|| Im \langle \phi_i|A_0|\phi_j \rangle \langle \phi_j|B_0|\phi_i \rangle | \}^2 \\
& \leq & \{ \sum_{i<j} 2 \sqrt{\alpha(1-\alpha)}|\lambda_i-\lambda_j| | \langle \phi_i|A_0|\phi_j \rangle || \langle \phi_j|B_0|\phi_i \rangle | \}^2 \\
& \leq & \{ \sum_{i<j} \{ (\lambda_i+\lambda_j)^2-(\lambda_i^{\alpha}\lambda_j^{1-\alpha}+\lambda_i^{1-\alpha}\lambda_j^{\alpha})^2 \}^{1/2} | \langle \phi_i|A_0|\phi_j \rangle || \langle \phi_j|B_0|\phi_i \rangle | \}^2 \\
& \leq & \sum_{i<j} (\lambda_i+\lambda_j-\lambda_i^{\alpha}\lambda_j^{1-\alpha}-\lambda_i^{1-\alpha}\lambda_j^{\alpha})| \langle \phi_i|A_0|\phi_j \rangle |^2 \\
&   &  \times \sum_{i<j} (\lambda_i+\lambda_j+\lambda_i^{\alpha}\lambda_j^{1-\alpha}+\lambda_i^{1-\alpha}\lambda_j^{\alpha})| \langle \phi_i|B_0|\phi_j \rangle |^2.
\end{eqnarray*}
Then we have 
$$
I_{\rho,\alpha}(A) J_{\rho,\alpha}(B) \geq \alpha(1-\alpha)|Tr[\rho [A,B]]|^2.
$$
We also have 
$$
I_{\rho,\alpha}(B) J_{\rho,\alpha}(A) \geq \alpha(1-\alpha)|Tr[\rho [A,B]]|^2.
$$
Hence we have the final result (\ref{eq:num3-1}). \ \hfill $\Box$ 

\begin{remark}
We remark that (\ref{eq:num2-3}) is derived by putting $\alpha = 1/2$ in (\ref{eq:num3-1}). Then Theorem \ref{th:theorem3-1} 
is a generalization of the result of Luo \cite{Lu:Hei}. 
\label{re:remark3-1}
\end{remark}

\begin{remark}
We remark that Conjecture 2.3 in \cite{FuYaKu:trace} does not hold in genaral. The Conjecture is (\ref{eq:num2-10}).  
A counterexample is given as follows. Let 
$$
\rho = \left( 
\begin{array}{cc}
3/4 & 0 \\
0 & 1/4 
\end{array} \right), \; A = \left( 
\begin{array}{cc}
0 & i \\
-i & 0
\end{array} \right), \; B = \left( 
\begin{array}{cc}
0 & 1 \\
1 & 0 
\end{array} \right), \; \alpha = 1/3.
$$
We have 
$$
I_{\rho,\alpha}(A) J_{\rho,\alpha}(B) = I_{\rho,\alpha}(B) I_{\rho,\alpha}(A) = 0.22457296 \cdots 
$$
and $|Tr[\rho [A,B]]|^2 = 1$. These imply 
$$
U_{\rho,\alpha}(A) U_{\rho,\alpha}(B) = 0.22457296 \cdots < \frac{1}{4}|Tr[\rho [A,B]]|^2 = 0.25.
$$
On the other hand we have 
$$
U_{\rho,\alpha}(A) U_{\rho,\alpha}(B) > \alpha (1-\alpha) |Tr[\rho [A,B]]|^2 = 0.2222222 \cdots.
$$
We also give a counterexample for Conjecture 2.10 in \cite{FuYaKu:trace}. The inequality 
$$
U_{\rho,\alpha}(A) U_{\rho,\alpha}(B) \geq \frac{1}{4}|Tr[(\frac{\rho^{\alpha}+\rho^{1-\alpha}}{2})^2 [A,B]]|^2
$$
is not correct in general, because $LHS = 0.22457296 \cdots, RHS = 0.23828105995 \cdots$. 
\label{re:remark3-2}
\end{remark}

\end{document}